\documentclass[fleqn,usenatbib]{mnras}

\usepackage[T1]{fontenc}

\DeclareRobustCommand{\VAN}[3]{#2}
\let\VANthebibliography\thebibliography
\def\thebibliography{\DeclareRobustCommand{\VAN}[3]{##3}\VANthebibliography}

\usepackage{graphicx}	
\usepackage{amsmath}	
\usepackage{amssymb}	
\usepackage[export]{adjustbox}
\usepackage{xcolor}
\usepackage{caption}
\usepackage{siunitx}

\newcommand{\de}{{\rm d}}

\usepackage{eso-pic}

\AddToShipoutPictureBG*{%
 \AtPageUpperLeft{%
  \hspace{0.65\paperwidth}%
  \raisebox{-4.5\baselineskip}{%
   \makebox[0pt][l]{\textnormal{FERMILAB-PUB-23-550-LDRD-PPD}}
}}}%

\title[Standard siren measurement of $H_0$ using O1-O3 GW events with DELVE]{A dark siren measurement of the Hubble constant using gravitational wave events from the first three LIGO/Virgo observing runs and DELVE}

\author[V. Alfradique \& Bom et al.]{
V. Alfradique$^{1}$\thanks{E-mail: vivianeapa@id.uff.br},
C. R. Bom$^{1,2}$,
A. Palmese$^{3}$,
G. Teixeira$^{1}$,
L. Santana-Silva$^{1}$,
A. Drlica-Wagner$^{4,5,6}$,
\newauthor
A. H. Riley$^{7}$,
C. E. Martínez-Vázquez$^{8}$,
D. J. Sand$^{9}$,
G. S. Stringfellow$^{10}$,
G. E. Medina$^{11}$,
\newauthor
J. A. Carballo-Bello$^{12}$,
Y. Choi$^{13}$,
J. Esteves$^{14}$,
G. Limberg$^{5,6,15}$,
B. Mutlu-Pakdil$^{16}$,
N. E. D. No\"el$^{17}$,
\newauthor
A. B. Pace$^{3}$,
J. D. Sakowska$^{18}$,
J. F. Wu$^{19,20}$
\\
\\
$^{1}$Centro Brasileiro de Pesquisas F\'isicas, Rua Dr. Xavier Sigaud 150, 22290-180 Rio de Janeiro, RJ, Brazil \\
$^{2}$Centro Federal de Educa\c{c}\~{a}o Tecnol\'{o}gica Celso Suckow da Fonseca,  Rodovia M\'{a}rcio Covas, lote J2, quadra J - Itagua\'{i} (Brazil)\\
$^{3}$McWilliams Center for Cosmology, Carnegie Mellon University, 5000 Forbes Ave, Pittsburgh, PA 15213, USA \\
$^{4}$Fermi National Accelerator Laboratory, P.O.\ Box 500, Batavia, IL 60510, USA \\
$^{5}$Kavli Institute for Cosmological Physics, University of Chicago, Chicago, IL 60637, USA \\
$^{6}$Department of Astronomy \& Astrophysics, University of Chicago, 5640 S Ellis Avenue, Chicago, IL 60637, USA\\
$^{7}$Institute for Computational Cosmology, Department of Physics, Durham University, South Road, Durham DH1 3LE, UK \\
$^{8}$Gemini Observatory/NSF's NOIRLab, 670 N. A'ohoku Place, Hilo, HI 96720, USA \\
$^{9}$Department of Astronomy/Steward Observatory, 933 North Cherry Avenue, Room N204, Tucson, AZ 85721-0065, USA \\
$^{10}$Center for Astrophysics and Space Astronomy, University of Colorado, 389 UCB, Boulder, CO 80309-0389, USA \\
$^{11}$Department of Astronomy and Astrophysics, University of Toronto, 50 St. George Street, Toronto ON, M5S 3H4, Canada \\
$^{12}$Instituto de Alta Investigación, Universidad de Tarapacá, Casilla 7D, Arica, Chile \\
$^{13}$NSF's NOIRLab, 950 N. Cherry Ave, Tucson, AZ 85719, USA \\
$^{14}$Department of Physics, University of Michigan, Ann Arbor, MI 48109, USA \\
$^{15}$Universidade de S\~ao Paulo, IAG, Departamento de Astronomia, SP 05508-090, S\~ao Paulo, Brazil \\
$^{16}$Department of Physics and Astronomy, Dartmouth College, Hanover, NH 03755, USA \\
$^{17}$Department of Physics, University of Surrey, Guildford GU2 7XH, UK \\
$^{18}$Department of Physics, University of Surrey, Guildford GU2 7XH, UK\\
$^{19}$Space Telescope Science Institute, 3700 San Martin Drive, Baltimore, MD 21218, USA \\
$^{20}$Center for Astrophysical Sciences, Johns Hopkins University, 3400 N. Charles St., Baltimore, MD 21217, USA 
}
\date{Accepted XXX. Received YYY; in original form ZZZ}

\pubyear{2023}

\begin{document}
\label{firstpage}
\pagerange{\pageref{firstpage}--\pageref{lastpage}}
\maketitle

\begin{abstract}
The current and next observation seasons will detect hundreds of gravitational waves (GWs) from compact binary systems coalescence at cosmological distances. When combined with independent electromagnetic measurements, the source redshift will be known, and we will be able to obtain precise measurements of the Hubble constant $H_0$ via the distance-redshift relation. However, most observed mergers are not expected to have electromagnetic counterparts,  which prevents a direct redshift measurement. In this scenario, one of the possibilities is to use the dark sirens method that statistically marginalizes over all the potential host galaxies within the GW location volume to provide a probabilistic redshift to the source. Here we presented $H_{0}$ measurements using two new dark sirens compared to previous analyses using DECam data, GW190924$\_$021846 and GW200202$\_$154313. The photometric redshifts of the possible host galaxies of these two events are acquired from the DECam Local Volume Exploration Survey (DELVE) carried out on the Blanco telescope at Cerro Tololo in Chile. The combination of the $H_0$ posterior from GW190924$\_$021846 and GW200202$\_$154313 together with the bright siren GW170817 leads to $H_{0} = 68.84^{+15.51}_{-7.74}\, \rm{km /s/Mpc}$. Including these two dark sirens improves the 68\% confidence interval (CI) by 7\% over GW170817 alone. This demonstrates that the inclusion of well-localized dark sirens in such analysis improves the precision with which cosmological measurements can be made. Using a sample containing 10 well-localized dark sirens observed during the third LIGO/Virgo observation run, we determine a measurement of $H_{0} = 76.00^{+17.64}_{-13.45}\, \rm{km /s/Mpc}$.
\end{abstract}

\begin{keywords}
catalogs -- cosmology: observations -- gravitational waves -- surveys
\end{keywords}



\section{Introduction}

The advent of gravitational wave measurements opened a new era of multi-messenger observation, shedding light on the properties of our universe. Standard sirens, a term introduced by \citet{schutz}, provide a way to measure cosmological parameters by restricting the distance-redshift relation. The gravitational wave detections provide a direct measure of luminosity distance without any additional distance calibrator, justifying the name "standard sirens" in analogy with standard candles. If a source has an electromagnetic counterpart, its redshift (\textit{z}) can be directly measured, and we referred to them as "bright standard sirens". The first bright standard siren measured was the binary neutron star (BNS) merger GW170817 \citep{ligobns}, whose electromagnetic gamma-ray burst counterpart was detected by the Fermi Gamma-ray Burst Monitor \citep{Goldstein_2017} and the anti-coincidence shield of the gamma-ray spectrometer on-board INTErnational Gamma-Ray Astrophysics Laboratory \citep{Savchenko_2017} within 0.1 to 0.647 seconds, and later complementing with the identification of the optical kilonova 
\citep[e.g.][]{Arcavi_2017, Coulter_2017, Cowperthwaite_2017, Soares_Santos_2017, Chornock_2017, Kasliwal_2017, Nicholl_2017, Evans_2017, Pian_2017, Smartt_2017, Tanvir_2017, Valenti2017} detected about 11 hours after the merger. This event produced the first direct and independent measure of $H_0$, $H_{0} = 70^{+12}_{-8}\,\rm{km/s/Mpc}$ \citep{2017Natur.551...85A}.

After a three year hiatus during which improvements in the sensitivity of the detectors were made, the upcoming fourth run of the LIGO, Virgo and KAGRA collaboration will be able to observe a larger fraction of the universe than previous observing runs and projected to detect an estimated $\sim$90 gravitational wave events per year with a $\sim$85\% improvement in sky localization of the source \citep{2018LRR....21....3A}. With more interferometers in operation (like the Einstein Telescope \citealt{Sathyaprakash_2012}, Cosmic Explorer \citealt{Abbott_2017} and the LISA space interferometer \citealt{lisa}), it is possible that in the next years more standard sirens will be identified which can lead to a $H_0$ measurement with precision in the same order as what is achieved with other cosmological probes such as the cosmological microwave background (CMB) \citep{planck18} and the Cepheid \citep{Riess_2021} or Red Giant Branch \citep{Freedman_2019}-calibrated type Ia supernovae. 
This new independent measurement of the Hubble constant can enable a way to clarify the origin of the observed current $4-6\sigma$ tension \citep{Valentino2021, Verde2019}. Despite these improvements, detection of the events electromagnetic counterparts remains a challenge, requiring dedicated follow-up campaigns and strategies \citep{bom2023designing}, particularly for those events involving black hole companions, which may have no electromagnetic signature emitted or be associated to a flare \citep{bom2023standard,rodriguez2023optical}.

A prime example of the challenge to localize and identify the electromagnetic emission of an event involving a black hole is the GW190814 event \citep{190814_paper}. GW190814 has an excellent sky localization (23 deg$^2$) and a high probability (>99\%) of being a black hole-neutron star merger,  and became an excellent candidate to provide the first detection of the counterpart of these types of systems. Several electromagnetic follow-ups, from gamma rays to radio, were started by different groups \citep[e.g.][]{Kilpatrick_2021, Tucker_2022, Dobie_2019, Gomez_2019, Ackley_2020, Andreoni_2020, Vieira_2020, Watson_2020, Alexander_2021, deWet_2021} with a continuous duration of up to more than 250 days after the merger. The properties of the electromagnetic counterpart candidates were analyzed and compared with the theoretical prediction for NSBH fusion, including optical spectra, variability of radio sources, their location, photometric evolution, and redshift of possible host galaxies. Despite immense dedicated effort, no sign of a gamma-ray burst or any optical counterpart has been identified, but allowed to discard some possible types of electromagnetic transients such as: kilonova with large ejecta mass $M\geq0.1M_{\odot}$ \citep{Ackley_2020}, "blue" kilonovae with $M>0.5M_{\odot}$ \citep{Kilpatrick_2021}, an AT2017gfo-like kilonova \citep{deWet_2021}, short gamma-ray burst with viewing angles less than 17$^\circ$ \citep{Kilpatrick_2021}, and a short gamma-ray burst-like Gaussian jet with a particular configuration \citep{Alexander_2021}. 
In view of these problems, an alternative to the lack of an electromagnetic counterpart is to use the redshifts of galaxies that are within the coalescence location volume to break the $H_{0}$-\textit{z} degeneracy and infer cosmological parameters. This methodology is known as \textit{dark standard sirens} (see \citealt{2023AJ....166...22G} for a review of the method).

The dark standard sirens approach was applied to constrain the cosmology in several LIGO and Virgo detections. \citet{fishbach} studied the event GW170807 and showed that the obtained precision of $H_0$ is about 3 times worse than the "bright" siren method \citep{2017Natur.551...85A}. \citet{darksiren1} and \citet{palmese20_sts} investigated the method with the Dark Energy Survey (DES) galaxy catalog for binary black hole (BBH) mergers (GW170814 and GW190814, respectively) and showed that a single dark siren BBH provides a measure of $H_0$ with a precision of $48\%$ for GW170814 and $55\%$ GW190814. Recently, \citet{Palmese_2023} demonstrated that 8 dark sirens well localized in the sky are able to provide a measurement as accurate as that obtained with a single bright siren GW170817 (about 20\% against 18\%; \citealt{2017Natur.551...85A}). 

\citet{chen17} predicted that 5 years of detections for LIGO, Virgo and KAGRA collaboration (at design sensitivity) could lead to a precision of $\sim$ 5\% and 10 \% of $H_0$ measurement for the BNS and BBH, respectively. For this result, they assumed that all events within 10,000 Mpc$^3$ will be detected and that complete galaxy catalogs will be available. In the next decade, the arrival of the next generation of terrestrial interferometers, such as the Einstein Telescope and the Cosmic Explorer, could rapidly increase the number of detections, allowing us to check the predictions of the percentage level of the measure of $H_0$ made by \citet{Muttoni2023}.

The intent of this study is to investigate the ability of the dark siren events GW190924$\_$021846 and GW200202$\_$154313 to constrain the Hubble constant. We combine our results with that of 8 dark sirens present in \citet{Palmese_2023} and perform the most precise $H_0$ measurement with the better localized dark sirens. The choice for these events is justified due to the small localization volume, which decreases the number of potential host galaxies to be marginalized over, and because their localization region is covered by DELVE\footnote{\url{https://delve-survey.github.io/}} \citep{Drlica_2021} galaxy catalogs. All the photometry redshift information is provided by the second release of DELVE data (DELVE DR2\footnote{\url{https://datalab.noirlab.edu/delve/}}, \citealt{Drlica_2022}), the galaxy photometric redshift was estimated using the Mixture Density Network (MDN, \citealt{Bishop_1994}), a machine learning technique that provides the probability density function (PDF) of the photo-\textit{z}. This technique uses magnitudes and color information to train the various Gaussian distributions that will be combined into the final PDF. In contrast to previous work, our work innovates by applying DELVE data to the standard siren methodology for the first time for two new events from the third observing run (O3), implementing a more refined artificial neural network technique for photo-\textit{z} measurements instead of the commonly used random forest algorithms \citep{galpro, zhou2020clustering}. The results of this study may provide insight into the potential of dark sirens as a cosmological probe, computing the precision level of $H_0$ measurement that this methodology can achieve with realistic photometric uncertainty and sky coverage.

This paper is organized as follows: in \S\ref{data} we describe the data used in the dark sirens methodology that is discussed in \S\ref{method}. Our results are presented and discussed in \S\ref{results}, and our final conclusions are presented in \S\ref{conclusions}. Throughout the article, we adopt a flat $\Lambda$CDM cosmology with $\Omega_m =0.3$ and $H_0$ values in the $20-140~{\rm km~s^{-1}~Mpc^{-1}}$ range. When not otherwise stated, quoted error bars represent the 68\% CI.

\section{Data}\label{data}

\subsection{LIGO and Virgo data: gravitational wave events}\label{sec:GWdata}

\begin{figure*}
\centering
\includegraphics[width=1\linewidth]{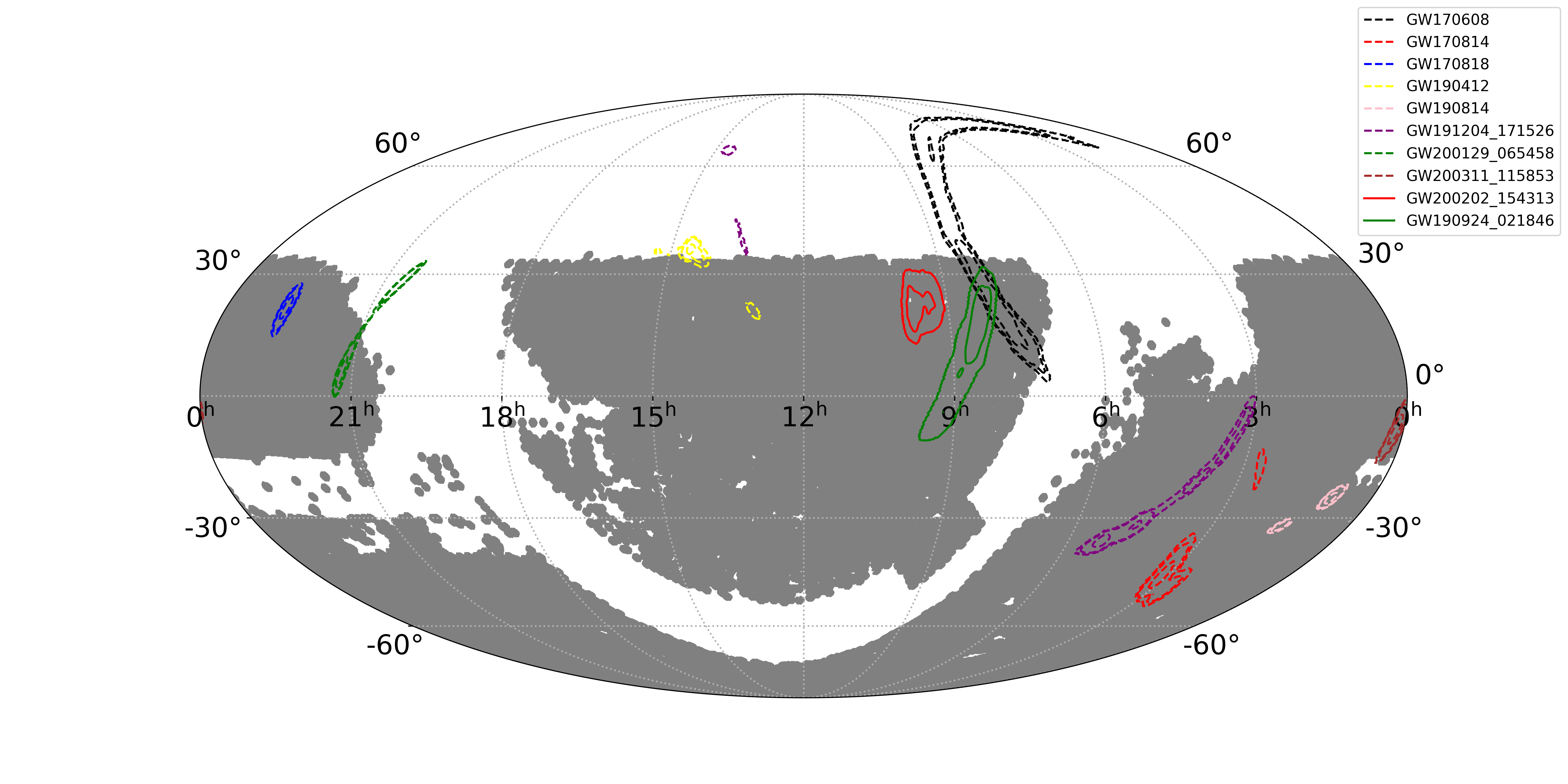}
\caption{LIGO/Virgo GW dark standard sirens analyzed in this paper, where the contours represent the 90\% CI localization from the sky maps. The shaded regions are those that are covered by the DELVE catalogs used in this work.}
\label{fig:data}
\end{figure*}

Here, we extend the 8-event catalog used in \citet{Palmese_2023} by adding two new events: GW190924$\_$021846 and GW200202$\_$154313. In total, our sample includes the 10 best localized events in the sky detected during the third LIGO/Virgo observing period. For these two added events, we used the gravitational wave data from the maps publicly available by the LIGO and Virgo collaboration in \citet{gwtc2} and \citet{gwtc3}. The right ascension (RA), declination (dec), and distance probability are given in HEALPIX pixels \citep{healpix}, where this probability is supposed to be Gaussian along each line of sight. GW200202$\_$154313 is the result of the merger of two black holes of approximately 7 and 10 solar masses, this is one of the best three-dimensional localizations from the second-half of the O3 (see Table \ref{tab:events}), having a 90\% credible volume of 0.0034 Gpc$^3$ and a 90\% CI area of 167 deg$^2$. The location of maximum probability is centered at RA = 146.25 deg and dec = 20.98 deg. Marginalizing over all other parameters, the estimate of the luminosity distance has a mean equal to 364.3 Mpc with a standard deviation of 90.2 Mpc. The second detection, GW190924$\_$021846 is the result of the merger of two lightest black holes \citep{gwtc2}, with inferred masses equal to $m_{1} = 8.9^{+7.0}_{-2.0} M_{\odot}$, $m_{2} = 5.0^{+1.4}_{-1.9} M_{\odot}$ and a 90\% credible volume of $\sim$0.02 Gpc$^3$ and a 90\% CI area of 348 deg$^2$. The GW190924$\_$021846 has a maximum probability of being located at RA, dec =$\left(134.561, 2.687\right)$ deg. At the peak location, the luminosity distance mean is 479.4 Mpc and the standard deviation is 151.7 Mpc. The area of the sky enclosing 90\% CI is 348 deg$^2$. Fig. \ref{fig:data} shows the 90\% CI contours of all events used in this work with the area covered by DELVE. The two new events are fully covered by DELVE, except for a small tip on GW$190924_021846$ containing approximately $2\%$.

\begin{table*}
\centering
\caption{Luminosity distance, 90\% CI area, and volume of gravitational wave events and candidates used in this analysis. We also report the reference paper or GCN that reports the sky map used for each event. These events have estimated false alarm rates of fewer than 1 in $10^{3}-10^{23}$ years. These candidates have all recently been confirmed as gravitational wave events in \citet{gwtc3}.
} 
\begin{tabular}{ccccc}
\hline
Event & $d_L$ [Mpc] & A [deg$^2$] & V [Gpc$^3$] & Reference\\
\hline
GW170608 & $320^{+120}_{-110}$ & 392 & $3 \times 10^{-3}$ &  \citet{gwtc1} \\
GW170814 & $540^{+130}_{-210}$ & 62 & $2 \times 10^{-3}$ &  \citet{gw170814}\\
GW170818 & $1060^{+420}_{-380}$ & 39 & $7 \times 10^{-3}$ &  \citet{gwtc1}\\
GW190412 & $740^{+120}_{-130}$ & 12 & $4 \times 10^{-4}$ &  \citet{GW190412}\\ 
GW190814 & $241^{+26}_{-26}$ & 19 & $3 \times 10^{-5}$ &  \citet{190814_paper}\\
S191204r & $678^{+149}_{-149}$ & 103 & $6 \times 10^{-3}$ &  \citet{191204}\\ 
S200129m & $755^{+194}_{-194}$ & 41 & $2 \times 10^{-3}$ &  \citet{200129}\\
S200311bg & $1115^{+175}_{-175}$ & 34 & $5 \times 10^{-3}$ &  \citet{200311}\\
GW200202\_154313 & $410^{+150}_{-160}$ & 167 & $3 \times 10^{-3}$ &  \citet{gwtc3}\\
GW190924\_021846 & $570\pm220$ & 348 & $2 \times 10^{-2}$ &  \citet{gwtc2}\\
\hline
\end{tabular}
\label{tab:events}
\end{table*}

\subsection{The galaxies photo-\textit{z}'s: a deep learning algorithm for DELVE data}\label{sec:delve}

The DELVE is a project that combines public data (including data from DES, Dark Energy Camera Legacy Survey and DECam eROSITA Survey) with more than 126 nights of novel observations made with the Dark Energy Camera (DECam, \citealt{flaugher}) located on the 4-meter Blanco Telescope at Cerro Tololo Interamerican Observatory in Chile. DELVE uses the combination of the large field of view (3 deg$^2$) and the fast readout time (27 s) of DECam with the expectation of providing complete coverage of the entire high-Galactic-latitude southern sky. In the WIDE observational program, the DELVE DR2 \citep{Drlica_2022} covers >17,000 deg$^2$ in the \textit{griz} bands out to 23.5 mag.

The photometric redshifts (or photo-\textit{z}s) for the DELVE data were computed using the deep learning method called Mixture Density Network. In brief, the method is a combination of a deep neural network with the assumption that any distribution can be written as a mixture of distributions (chosen to be the normal distribution in its traditional form). The deep neural network is trained, given some input features, to select the best parameters of the multiple distributions that will be mixed into a single distribution. The output parameters used are the mean, standard deviation, and mixing coefficients, which are the probabilistic weights of each normal distribution. In this way, the MDN is capable to reproduce the galaxy photo-\textit{z} PDF, given some input features. The input features are the \textit{griz} magnitudes, and the \textit{g-r}, \textit{g-i}, \textit{g-z}, \textit{r-i}, \textit{r-z}, \textit{i-z} colors. In the next sections, we use this approach to compute the photo-\textit{z}s of the possible galaxy host whenever the spectroscopic redshift is not available.  

The MDN was implemented with the following structure: a LMU layer with 212 units; a 2-layer Multi-Layer Perceptron with 86 units each; a Dropout layer with 20\% rate; and finally a \texttt{MixtureNormal} layer that returns the outputs (the mean, standard-deviation and weights of the 20 Gaussian distributions). The LMU layer was implemented using the \texttt{keras-lmu} \citet{voelker2019lmu} application; the inner Perceptron and Dropout layers, the standard DL framework and the \texttt{MixtureNormal} output layer were built within the \texttt{tensorflow}  and \texttt{tensorflow-probability} libraries API\footnote{Tensorflow v2.9.1; Tensorflow Probability v0.17.0; keras-lmu v0.5.0} \citep{tensorflow2015-whitepaper}.The architecture of the network also incorporates a Legendre Memory Unit \citep[LMU,][]{lmu_cell} Layer at the head of the network. This architecture was one of the networks submitted in the LSST-DESC Tomography Optimization Challenge \citep{Zuntz_2021}, and it exhibited the best performance for the DELVE DR2 photo-\textit{z}'s regression task. We combined the photo-\textit{z} PDF estimated by the MDN output layer \citep[also used for photometric redshift regression in the S-PLUS Survey in][]{Lima_2022} with the well-performing LMU layer to estimate the photometric redshifts. More details can be found in Teixeira et. al. (in preparation).
Following the work of \citet{Zuntz_2021}, the LMU layer is included to more efficiently assign galaxies to redshift bins, selecting relevant information from previous data while simultaneously removing expendable data. For the loss function, we chose the maximum likelihood, which was minimized with the Nadam Optimizer \citep{Dozat_2016} and results in a learning rate of 0.0002.

The network was trained to maximize the PDF peak value for the spectroscopic redshifts ($z_{spec}$) of each galaxy. 
The spectroscopic information came from a crossmatch between DELVE DR2 and the data available in different large sky surveys \citep{Ahumada_2020, Colless_2001, Jones_2009, refId0, Baldry_2014, Mortlock_2001, Newman_2020, Drinkwater_2010, Newman_2013, Momcheva_2016, Le_F_vre_2013, Mercurio_2021, Cooper_2012, Holwerda_2011, Mao_2021, Bayliss_2016, Bradshaw_2013, McLure_2012, Masters_2019, Mao_2012, Bacon_2010, McLure_2018, wilson2006clusters, Treu_2015, Pharo_2020, Tasca_2017, Wirth_2015, Nanayakkara_2016}, which resulted in approximately 4.5 million galaxies with $z_{spec}$ measurements. We also added the $z_{spec}$'s available from the DECals DR9 Catalog \citep{Dey_2019} also by doing a crossmatch with the DELVE DR2 data. All the matches were made considering a maximal separation of $0.972$ arcseconds.

In order to guarantee high quality photometric data used to train and test the model, we apply the following constraints on the colors, signal-to-noise ratio (SNR), and the limit of $z_{spec}$:
\begin{itemize}
\centering
    \item[] SNR > 3 for \textit{g} 
    \item[] SNR > 5 for \textit{riz}
    \item[] $-1<g-r<4$
    \item[] $-1<r-i<4$
    \item[] $-1<i-z<4$
    \item[] $g<22.5$   
    \item[] $0.01<z_{spec}<1$
\end{itemize}

The SNR cuts were used to eliminate spurious sources, bad measurements, and very faint galaxies. The $g$ mag limitation serves to reinforce the exclusion of faint galaxies. The color cuts were made in order to eliminate nonphysical (extremely blue and extremely red) objects (see \citealt{Drlica_Wagner_2018}), thus the majority of the objects in our sample populate the color-color diagram in the regions $-0.5 \leqslant r -i \leqslant 1.5$ and $-0.5 \leqslant r -i \leqslant 0.8$. We restricted our $z_{spec}$ interval to avoid spurious detections of low  surface brightness galaxies located at high redshift. We also used the \texttt{MODEST\_CLASS} criteria \citep{Drlica_Wagner_2018} to remove contaminant stars by choosing the objects that lie in the classes $1$ (high-probably galaxy) and $3$ (ambiguous classification).

To account for the lack of a band on our data, we decided to train 3 different MDN's for each different observation scenario: (1) full coverage, with optical data in \textit{griz} bands and partial coverage when we are missing a band - coverage only in (2) \textit{gri} bands or (3) \textit{grz} bands. Each MDN was trained with the magnitude (and colors) appropriate to the different scenarios. We used all of them to predict the $z_{phot}$'s. The objects with full coverage were assigned to the flag \texttt{model\_GRIZ}, and the same was made for \textit{gri} and \textit{grz} coverages with the flags \texttt{model\_GRI} and \texttt{model\_GRZ}, respectively. For example, on the GW200202$\_$154313 event we have approximately 3.4M objects with estimated $z_{phot}$'s, being $\sim62\%$, $\sim30\%$ and $\sim8\%$ of the objects covered by \textit{griz}, \textit{grz}, \textit{gri} bands, respectively.

\begin{figure*}
\centering
\includegraphics[width=0.49\linewidth]{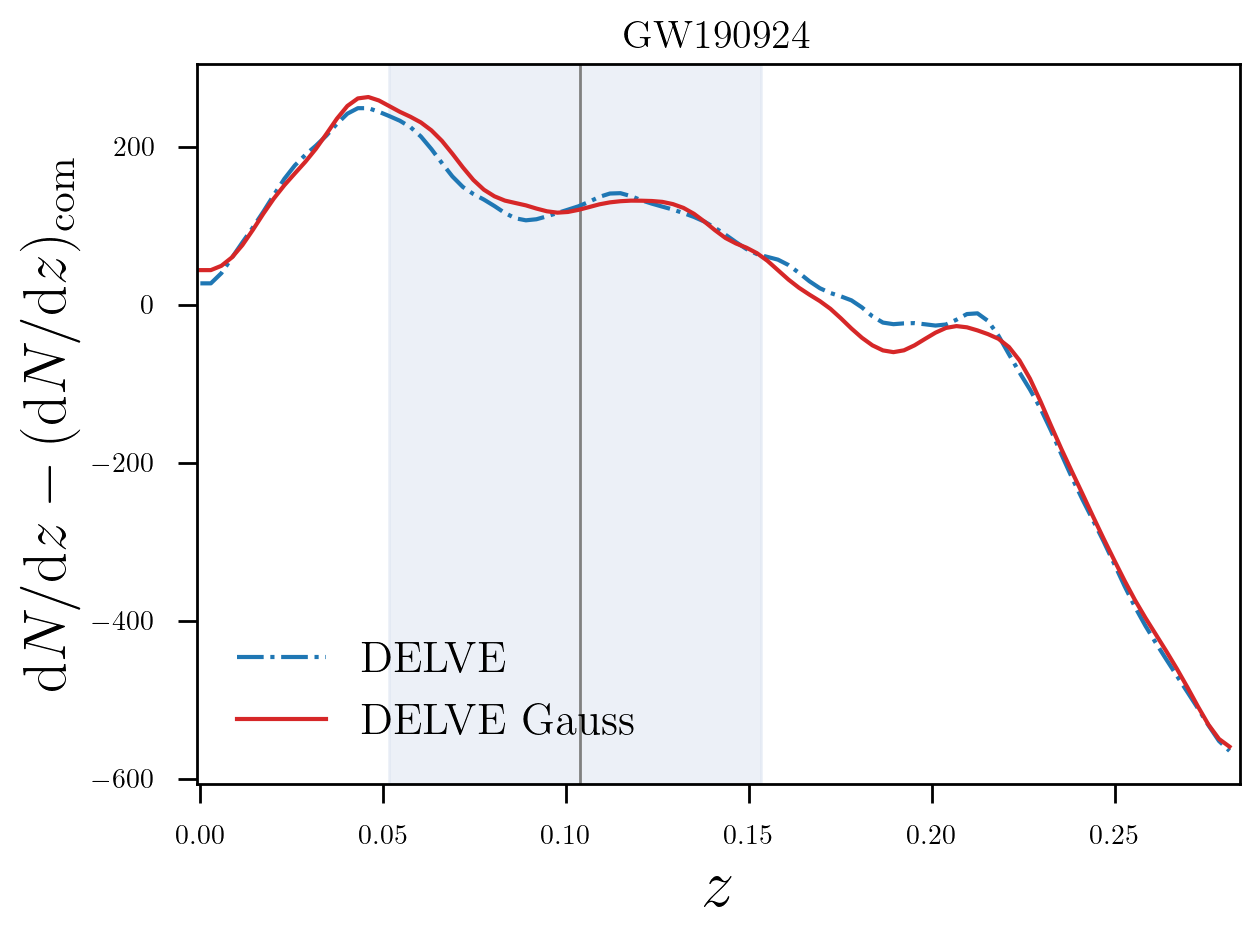}
\includegraphics[width=0.49\linewidth]{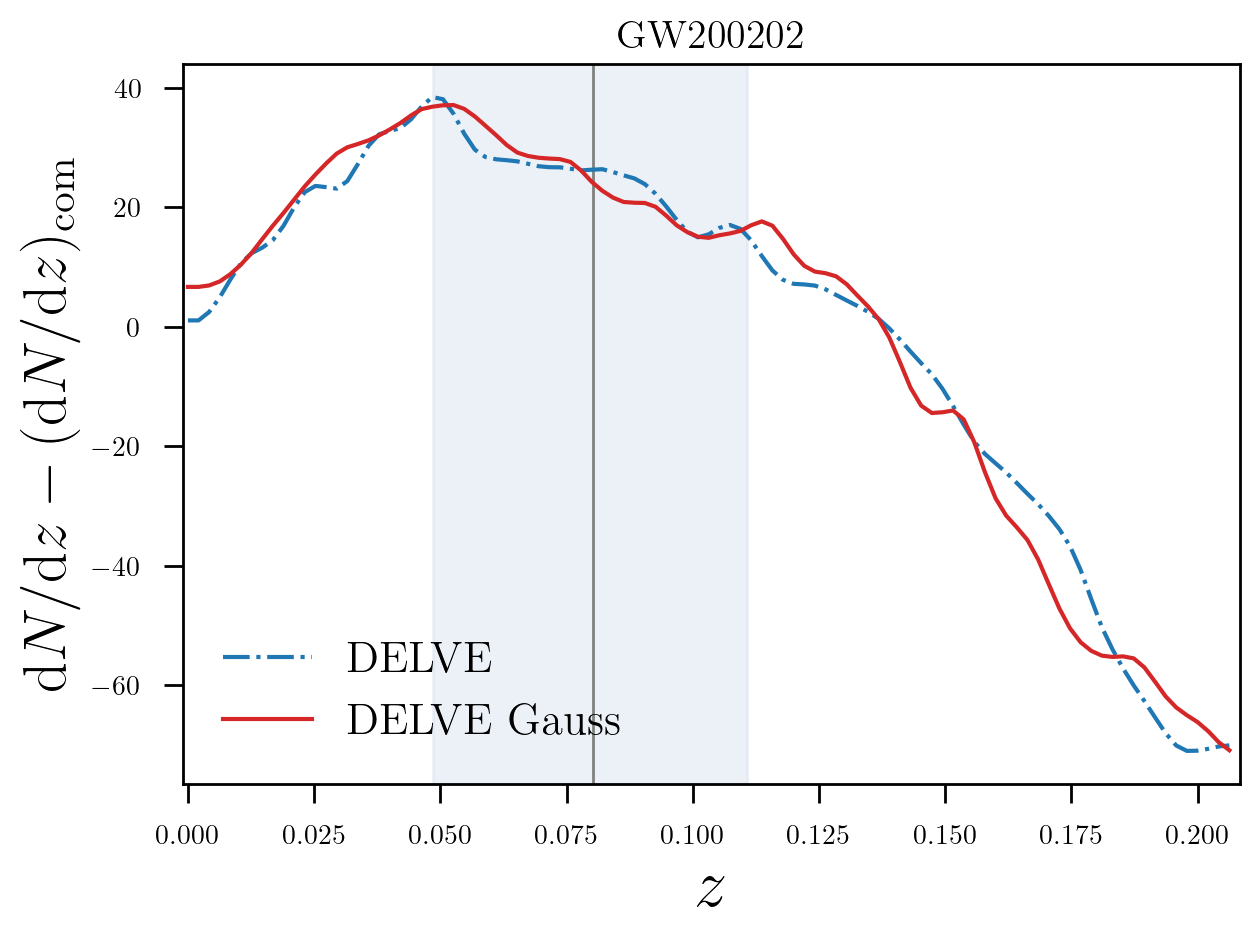}
\caption{Redshift distribution of galaxies in the 90\% CI area of the dark siren events analyzed in this work. The distribution is subtracted with a $dN/dz$ with uniform number density to highlight the presence of overdensities and underdensities along the line of sight. The dashed blue line shows the distribution using the photometric redshift point estimates from the DELVE, the red line shows the same redshifts when their uncertainty is considered as a Gaussian error. The gray vertical lines represent the luminosity distance of each GW event marginalized over the entire sky, assuming an $H_0$ of 70 km/s/Mpc, and the shaded regions are the $1\sigma$ uncertainties considering the same $H_0$; these regions are only shown for reference.}
\label{fig:dndz}
\end{figure*}

After the selection cuts, our training sample contains about one million objects distributed at redshift \textit{z} < 1. There are 31252 and 6367 of these galaxies in the 90\% probability region of GW190924$\_$021846 and GW200202$\_$154313, respectively. Fig.~\ref{fig:dndz} shows this final distribution, where we plotted the redshift distribution d\textit{N}/d\textit{z} subtracted from uniform number density (d\textit{N}/d\textit{z})$_{\rm{com}}$ assuming $H_{0} = 70\, \rm{km/s/Mpc}$ to emphasize the presence of overdensities along the line of sight.

\begin{figure*}
\centering
\includegraphics[width=0.49\linewidth,valign=T]{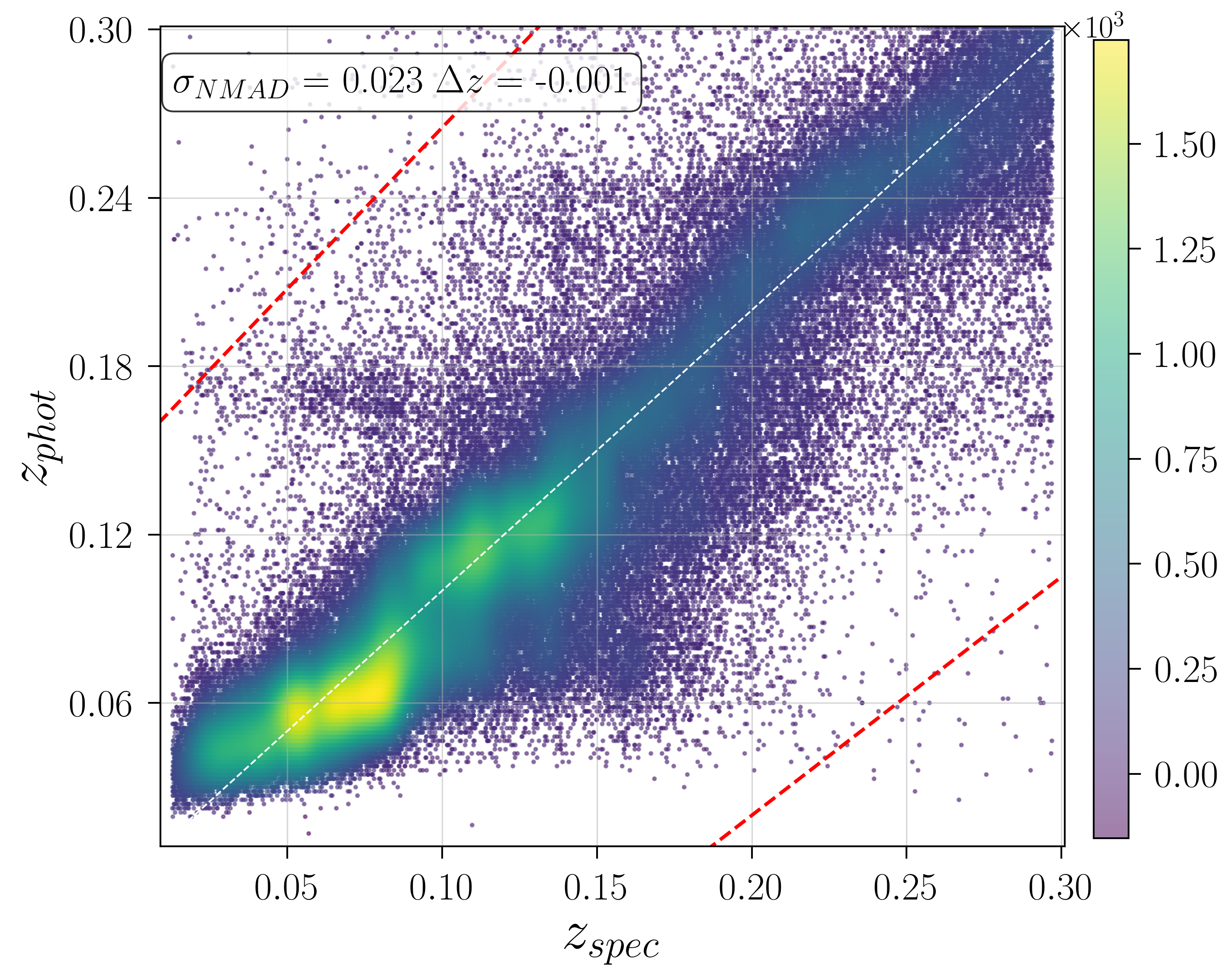}
\includegraphics[width=0.491\linewidth,valign=T]{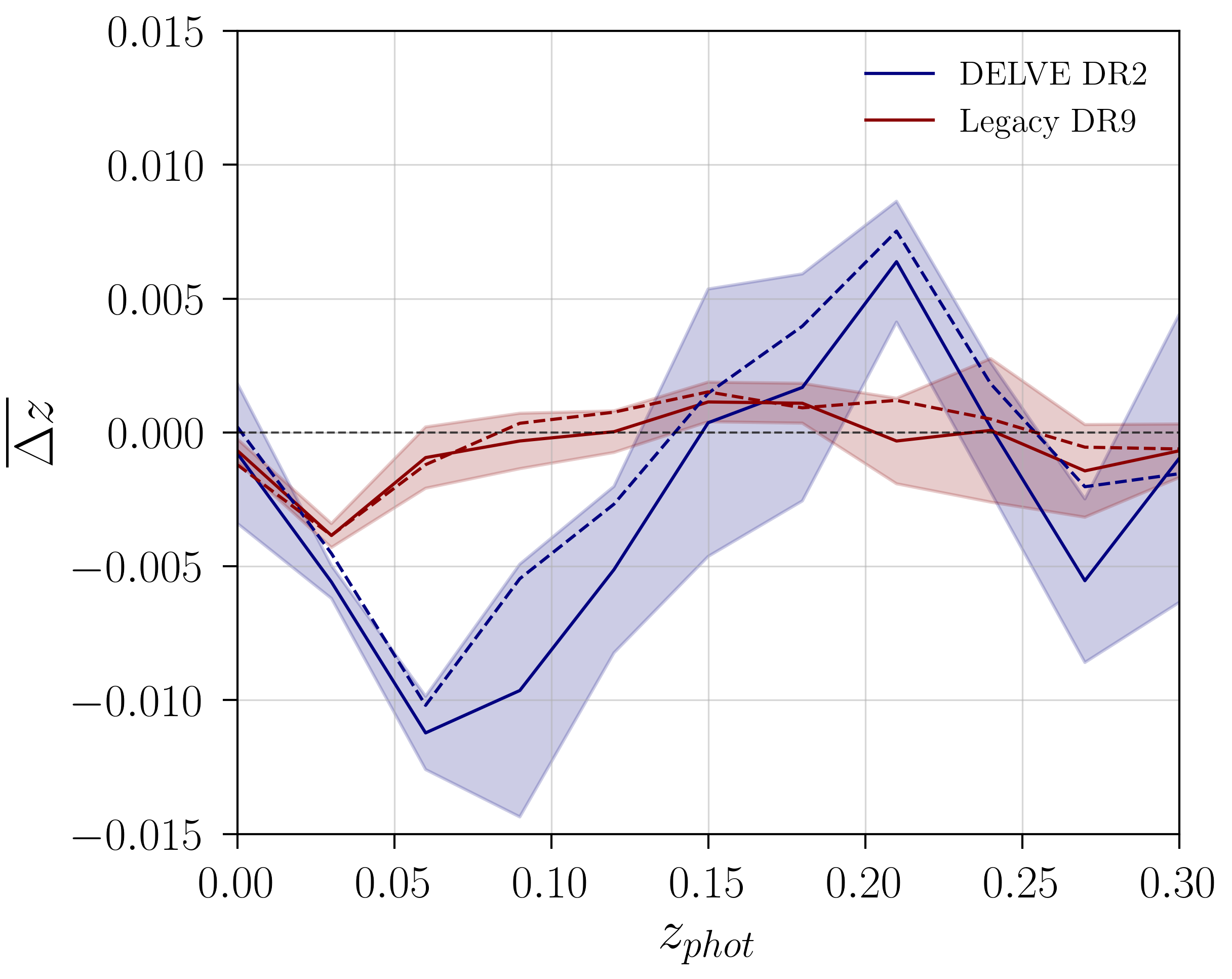}
\caption{Photometric redshift quality assessment plots using the testing sample from the available spectroscopic data. 
\emph{Left:} density
plot of galaxies in the validation sample, showing the predicted $z_{phot}$'s (PDF peaks) as a function of spectroscopic redshift. The red dotted lines represent the outliers limits, where outliers $z_{phot}^{out}$ are  defined as $|z_{phot}^{out} - z_{spec}| > 0.15/(1+z_{spec})$.
\emph{Right:} median value of the bias distribution $\Delta z = z_{phot} - z_{spec}$ in bins of photometric redshift for our model used in the DELVE DR2 and the Legacy Survey DR9 photometric redshifts. The median was calculated taking 3 different non-correlated regions in the sky covered by our test sample. The regions were chosen to have the same area as GW200202$\_$154313.
Both panels are plotted with $r<21\;mag$, $z_{spec}<0.3$ and $z_{phot}<0.3$.}
\label{fig:data2}
\end{figure*}

To evaluate the performance of our MDN method, we performed a complete analysis, evaluating the point statistics and PDF's metric for the validation sample (represented by the 2.3$\times 10^{5}\, z_{spec}$'s that have not been used for training the photo–z’s). The predicted photo-\textit{z}'s as a function of the measured spectroscopic redshifts is shown in the left panel of Fig.~\ref{fig:data2}. We can see that the majority of data points lie close to the diagonal, thus pointing to the accuracy of the predicted redshifts. Additionally, we can see the presence of outliers in every redshift interval. However, the outlier fraction (which is defined as $|\Delta z|>0.15\times\left(1+z_{spec}\right)$) results indicate that these data points only represent a minimum fraction (< 4\%) of the entire sample over the redshift range of interest. In order to avoid any systematic biases in DELVE galaxy distribution and their photo-\textit{z}, we select three different areas with the same size of LIGO 90$\%$ probability region and analyze the photo-\textit{z} quality for these regions (solid lines and shadows in the right panel in Fig. \ref{fig:data2}). The right panel of Fig. \ref{fig:data2} shows the median photo-\textit{z} bias in photo-\textit{z} bins of size 0.025 for DELVE and LEGACY-DR9 measurements. The results for DELVE full spectroscopic sample (dashed red line) and DELVE limited areas (dashed red line) revealed that the photo-\textit{z} bias is under control at $z_{\rm phot} = 0.5$, having median bias values smaller than 0.01 for each photo-\textit{z} bins and when considering the complete sample, the value reduces to -0.001. Thus, the measurements are uniform over the DELVE footprint. In contrast, the photo-\textit{z} results from the LEGACY full spectroscopic sample (blue dashed line) appear to outperform DELVE, displaying median bias values consistently below 0.005. This difference in quality could be attributed to the fact that LEGACY measurements benefit from uniform coverage across all bands and also leverage the advantages of infrared bands in their Spectral Energy Distributions (SEDs).
The scatter of $z_{phot}$ predictions was quantified with the normalized median absolute deviation, defined as $\sigma_{\rm{NMAD}} = 1.48\times\rm{median}\left(|\Delta z|/\left(1+z_{spec}\right)\right)$, and the 68th percentile width of the bias distribution about the median ($\sigma_{68}$). Data for DELVE objects brighter than r < 21 yields $\sigma_{\rm{NMAD}} = 0.023$ for all the galaxies in $0< z_{spec} <0.3$ and the $\sigma_{68}$ is less than 0.04 for all the photo-\textit{z} bins. These results are in agreement with previous works that use similar techniques to measure photometric redshift in large sky surveys \cite[see,][]{Lima_2022}. In order to validate the individual photo-\textit{z} PDF as a whole, we use two different metrics: the probability integral transform (PIT) distribution and the Odds value. The PIT distribution for DELVE data has a positive skewness, which indicates that our deep learning methods overestimate the $z_{phot}$. The Odds value represents the fraction of the photo-\textit{z} PDF that is contained in the interval $z_{spec}\pm0.06$, its distribution reveals that our MDN models produce narrow photo-\textit{z} PDFs with their values centered near 1. 

As shown previously, \citep{darksiren1, palmese20_sts, Palmese_2023} to overcome the fact that our sample is magnitude-limited, we have to ensure that it is volume-limited. For that, we follow the same steps used in \citet{Palmese_2023}. For each GW event, we start by computing the maximum redshift after converting the higher $90\%$ CI bounds in luminosity distance into the redshift, adopting the largest value of $H_0$ we considered in the prior. The next step is to find an absolute magnitude threshold value that corresponds to the apparent magnitude limit at the maximum redshift. Finally, we exclude all galaxies in our sample that have an absolute magnitude above this threshold (-19.39 and -20.32 for GW200202$\_$154313 and GW190924$\_$021846, respectively).

\section{Method}\label{method}

In this work, we used the Bayesian formalism, described in detail in \citet{chen17} and adapted into \citep{darksiren1, palmese20_sts, Palmese_2023}, to estimate the posterior probability of $H_0$ for the dark siren method. The $H_0$ posterior for a gravitational wave measurement $d_{\rm GW}$ and electromagnetic data $d_{\rm EM}$ for a galaxy survey is written via Bayes' theorem as
\begin{equation}
p\left(H_{0}|d_{\rm GW}, d_{\rm EM}\right) \propto p\left(d_{\rm GW},d_{\rm EM}|H_{0}\right)p\left(H_{0}\right),
\label{eq:posteriormain}
\end{equation}
where $p\left(H_{0}\right)$ is the prior on $H_0$ and $p\left(d_{\rm GW},d_{\rm EM}|H_{0}\right)$ is the joint GW–EM likelihood. Assuming that the GW and EM measurements are independent, the joint likelihood can be written as $p\left(d_{\rm GW},d_{\rm EM}|H_{0}\right) = p\left(d_{\rm GW}|H_{0}\right)p\left(d_{\rm EM}|H_{0}\right)$. By marginalizing over the true redshift, the sky position of the GW source, the photo-\textit{z} bias $\Delta z$ and over all the possible galaxy hosts, the $H_{0}$ posterior can be written as in \citet{Palmese_2023}:
\begin{align}
\begin{split}
p\left(H_{0}|d_{\rm GW}, d_{\rm EM}\right) \propto \frac{p\left(H_{0}\right)}{\beta \left(H_{0}\right)} \sum_i \frac{1}{\mathcal{Z}_i}\int p(d_{\rm GW}|d_{L}\left(z,H_{0}\right),\hat{\Omega}_i) \\
\times p_i(d_{\rm EM}|z,\Delta z) p\left(\Delta z\right) 
\frac{r^2\left(z\right)}{H\left(z\right)}\de z\de \Delta z, 
\end{split}
\label{eq:like3}
\end{align}
where $r\left(z, H_{0}\right)$ is the comoving distance, $H\left(z\right)=H_{0}\left(\Omega_{m}\left(1+z\right)^3+1-\Omega_{m}\right)^{1/2}$ is the Hubble parameter in a Flat $\Lambda$CDM model, $p\left(\Delta z\right)$ is the prior on the photometric redshift bias, $\beta\left(H_{0}\right)$ is the selection function responsible for normalizing the likelihood, and $\mathcal{Z}$ is the evidence term defined as $\mathcal{Z}=\int dz_{i} p\left(d_{\rm EM}|z_{i}\right)r^2\left(z_{i}\right)/H\left(z_{i}\right)$. The above expression includes the assumption that the source of the GW is located in one of the galaxies present in the galaxy catalog, making it a function of the solid angle $\hat{\Omega}_{i}$ and the redshift of each galaxy.

The above posterior has two important ingredients: the selection effect defined by the $\beta$ function and the photo-\textit{z} bias $\Delta z$. The first is associated with the selection effects adopted in the measurement process (of the electromagnetic counterparts and the detection of gravitational waves), the $\beta\left(H_{0}\right)$ function is computed following the same steps described in \citet{chen17} and \citet{Palmese_2023}.
For the electromagnetic emission selection effects, we used galaxies from the DELVE DR2 catalog distributed up to the known absolute magnitudes for each of the GW events in the analysis, where we consider only those in $z<0.5$. As reported in \citet{chen17} despite this being a simplification of the real EM selection effect, since it disregards any sky accessibility, weather, and observing conditions, it is still a coherent approximation for estimating the observation of the real-time electromagnetic follow-up. On a large scale, we assume that galaxies are isotropically distributed across the sky. By marginalizing over the entire sky, the selection function can be written as
\begin{equation}
    \beta\left(H_{0}\right) = \int p^{\rm GW}_{\rm sel}\left(d_{L}\left(z,H_{0}\right)\right)p\left(z\right)dz,
\end{equation}
where $p\left(z\right)$ is the distribution of possible host galaxies and $p^{\rm GW}_{\rm sel}\left(d_{L}\left(z,H_{0}\right)\right)$ is the probability of a source located at $d_{L}$ being detected. This term quantifies the GW selection effect introduced by detector sensitivity and detection conditions. For the computation of $\beta\left(H_{0}\right)$ we follow the same steps as \citet{Palmese_2023}: first we simulate 70,000 BBH mergers for 20 different values of $H_{0}$ within our prior range $\left[20,140\right] \rm{km/s/Mpc}$. The BBH population is distributed through the redshift distribution $p\left(z\right)$ which is a function of the merger rate  evolution and the cosmology-dependent comoving volume element. For simplicity, we assume that the merger rate follow the Madau-Dickinson star formation rate \citep{Madau_2014}. The mass of the black holes is distributed according to a power-law with index 1.6 (in agreement with the results found in \citealt{Properties_O3a}). We draw spins from a uniform distribution between $\left(-1,1\right)$. The GW signals were generated using the BAYESTAR software \citep{bayestar, Singer_2016, Singer_supp} using the frequency domain approximant IMRPhenomD. Finally, we assume the O3 sensitivity curves for LIGO and Virgo\footnote{Available at \url{https://dcc.ligo.org/LIGO-P1200087/public}}, use a matched-filter analysis, and calculate the SNR of each event. We assume, as a detection condition, that the network SNR is above 12 and at least 2 detectors have a single–detector SNR above 4.

Another important effect considered in our analysis is the photo-\textit{z} bias correction. When we are dealing with simulated data, the machine learning algorithm used for photometric redshift estimates can provide a biased redshift probability distribution function. The non-uniform training samples can cause systematic biases in the photo-\textit{z}, causing the peak of the distribution to be shifted by $\Delta z$ from the true value of \textit{z}. In order to consider this effect, we use the photo-\textit{z} bias computation\footnote{\url{https://datalab.noirlab.edu/delve/photoz.php}} for the DELVE DR2 catalog (see the detailed description of this calculation in section \ref{sec:delve}) in different values of \textit{z} that enter on $H_0$ posterior through $p\left(d_{\rm EM}|z, \Delta z\right)$. 

This methodology can be extended to a sample of multiple events \textit{j} with a combined data $\{d^{j}_{\rm GW}, d^{j}_{\rm EM}\}$, if we assume that the events are independent of each other and that they share the same galaxy catalog. The Hubble constant posterior can be written as the product of the single event \textit{j} likelihoods:
\\
\begin{equation}
    p\left(H_{0}|\{d_{{\rm{GW}},i}\}, d_{\rm EM}\right) \propto p\left(H_{0}\right)p\left(d_{\rm EM}|H_{0}\right)\prod_{j}p\left(d_{{\rm{GW}},j}|H_{0}\right)
\end{equation}

\section{Results and Discussion} \label{results}

\begin{figure*}
    \centering
    \includegraphics[width=0.8\linewidth]{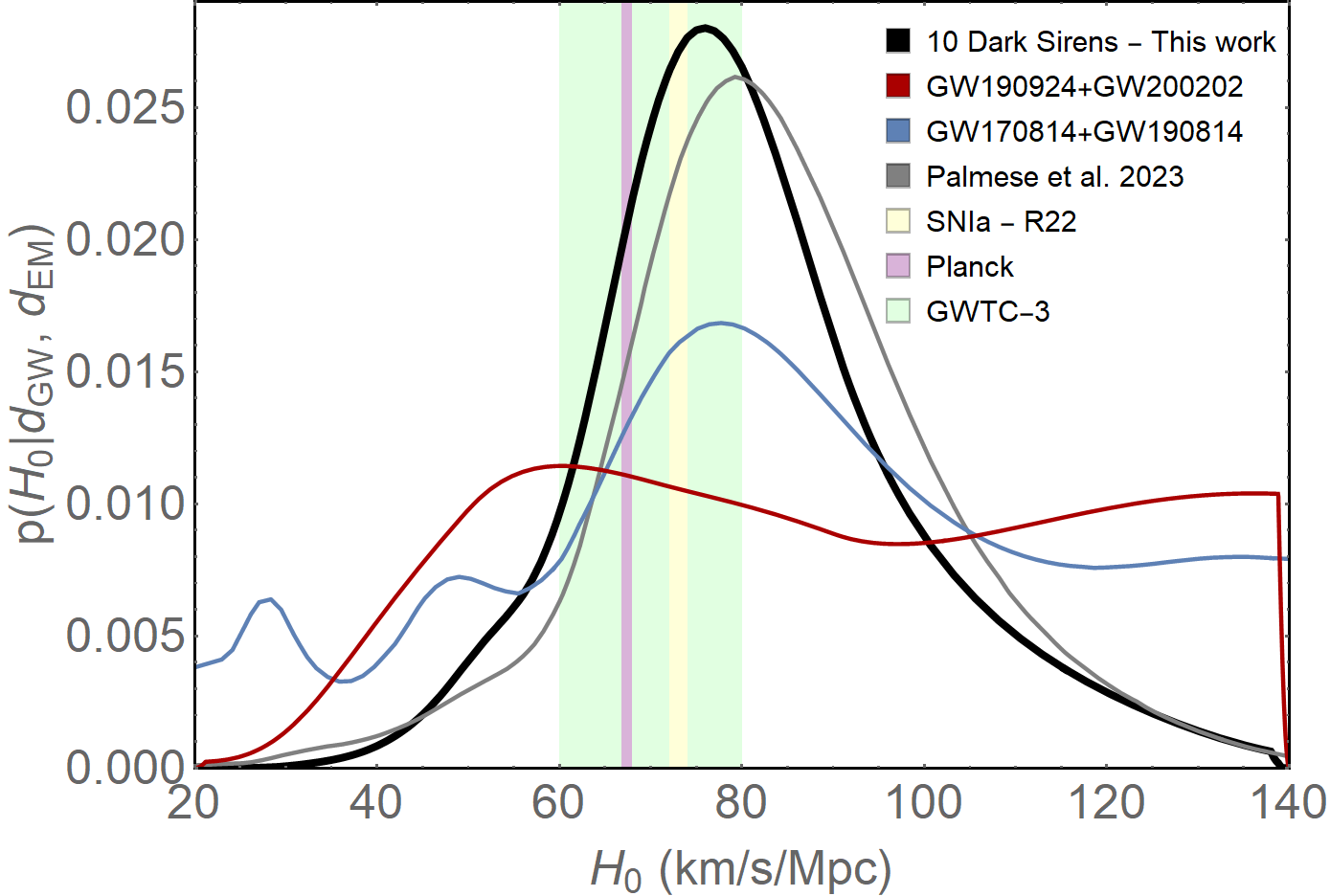}
    \caption{Hubble constant posterior distributions for the dark sirens considered in this work and previous works. The black line is the result of the combination of the two dark sirens considered here with the posteriors of all the 8 dark sirens (gray line) found in \citet{Palmese_2023}. The combination $H_0$ posteriors of GW190924$\_$021846 and GW200202$\_$154313 is shown in red, and for comparison we also show the combination result of GW170814 and GW190814 (blue line) presented in \citet{palmese20_sts}. For comparison, we show the 1$\sigma$ constraints on $H_0$ found by \citet{planck18}, \citet{Riess_2021} (R22) and \citet{Abbott_2023} (GWTC-3) as the vertical shaded regions.}
    \label{fig:posteriors}
\end{figure*}

\begin{figure*}
    \centering
    \includegraphics[width=0.8\linewidth]{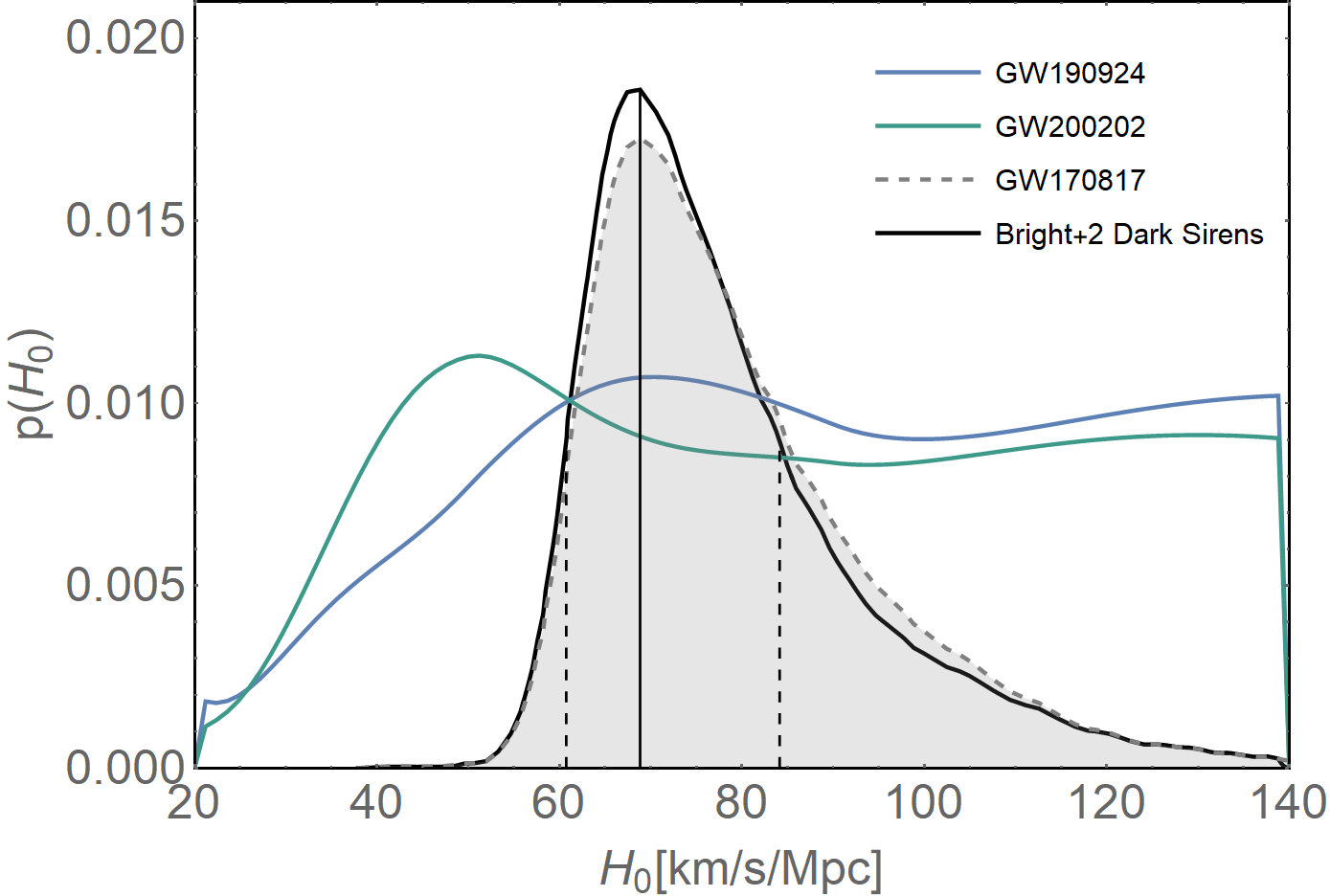}
    \caption{Hubble constant posterior distributions found using the DELVE galaxies for GW200202$\_$154313 (green line) and GW190924$\_$021846 (blue line). The dashed line represents the GW170817 bright siren result adapted from \citet{nicolaou2019impact}, which includes the peculiar velocity corrections for the galaxy host NGC 4993. The  black line is the result from the combination of the two dark sirens from this work with GW170817, and the vertical dashed lines show the 68\% region for this posterior. Posteriors are arbitrarily rescaled only for visualization purposes.}
    \label{fig:result}
\end{figure*}

We now use the DELVE photo-\textit{z}'s in the dark siren methodology to produce the $H_0$ posterior for GW190924$\_$021846 and GW200202$\_$154313. Then we combine the results for these two new GW events with those for 8 dark siren events (GW170608, GW170818, GW190412, S191204r, S200129m and S200311bg, GW170814 and GW190814) from \citep{palmese20_sts, Palmese_2023}. The first five events were found in \citet{Palmese_2023} using the DESI Legacy Survey galaxies’ redshifts, and the last two are presented in \citet{palmese20_sts} with the photo-\textit{z} catalog from DES. Fig.~\ref{fig:posteriors} shows the $H_{0}$ posterior from the combination of these two new dark sirens (dark red curve) and the final result (black curve) after combining the posterior of all the ten dark siren events. For comparison, we also show the results (blue curve) found in \citet{palmese20_sts} with the dark sirens GW170814 and GW190814 and for the 8 well-localized events (dark gray curve) found in \citet{Palmese_2023}. The two new events reduce the 68\% CI of the $H_0$ prior to values close to those found in \citet{Palmese_2023} (see Table \ref{tab:H0results}): GW190924$\_$021846 is able to reach the value of 85\% and GW200202$\_$154313 achieve the constraint of 90\%. The $H_0$ posterior distributions for GW190924$\_$021846 and GW200202$\_$154313 are presented in Fig.~\ref{fig:result}, we can see that both dark sirens display an evident peak at a low value of $H_0$ (near 70 km/s/Mpc for GW190924$\_$021846 and $\sim$51 km/s/Mpc for GW200202$\_$154313) that is a consequence of the notable overdensity of galaxies (see Fig. \ref{fig:dndz}) around redshift 0.05 to 0.1 and 0.05 for GW190924$\_$021846 and GW200202$\_$154313, respectively. As a result of better localization volume, which corresponds to a marginalization over a smaller number of galaxies, we can see that the posterior of GW200202$\_$154313 has a narrower peak, but the presence of a secondary peak at $H_0\sim 114$ km/s/Mpc makes it flatter than GW190924$\_$021846 (its kurtosis value is lower, $\sim$1.79, than that produced by GW190924$\_$021846, $\sim$1.95). The analysis of the skewness showed that event GW190924$\_$021846 produces a slightly more asymmetric posterior (a relative difference of approximately 58\%) than GW200202$\_$154313. The individual posteriors shown in Fig.~\ref{fig:result} present a high probability at the high $H_0$ end. The same result was observed for the $H_0$ posterior of GW190814 and GW170814 in \citet{palmese20_sts}, as explained by the authors, this is a consequence of the fact that the GW analysis only provides a $d_L$ estimate that can correspond to either a low \textit{z} and $H_0$ or a high \textit{z} and $H_0$. The cut on the prior range does not bias the final result, as it only changes the redshift range considered in the dark siren analysis and not the posterior behavior.

The combined result of all 10 dark sirens is shown in black in Fig. \ref{fig:posteriors}. The mode of the final posterior and the 68\% CI is $H_{0}=76.00^{+17.64}_{-13.45}\,\rm{km/s/Mpc}$. The addition of the two new dark sirens causes a reduction of $\sim$1\% over the 68\% confidence region found in \citet{Palmese_2023}. Table \ref{tab:H0results} summarizes our results and compares the performance with other standard siren analyses.

\begin{table*}
    \centering
    \caption{Hubble constant measurements found with the dark sirens from the three LIGO/Virgo runs and the bright siren GW170817. All priors are flat in the range $\left[20,140\right]$. The uncertainty from the flat prior is derived by assuming the same $H_0$ maximum found in the analysis. Quoted uncertainties represent 68\% HDI around the maximum of the posterior. The "$\sigma_{H_0}/\sigma_{\rm prior}$" column shows the 68\% CI from the posterior divided by 68\% CI of the prior width.}
 	\setlength{\tabcolsep}{2.2pt}
 	\begin{tabular}{l c c c c c}
 		\hline
 		Event  & $H_0\,\left(\rm{km/s/Mpc}\right)$ & $\sigma_{H_0}\,\left(\rm{km/s/Mpc}\right)$ & $\sigma_{H_0}/\sigma_{\rm prior}$ & Reference \\
            \hline
 	GW190924$\_$021846 & 
        $70.4^{+54.7}_{-15.1}$ & 34.9 & 85\% & This work  \\
 	GW200202$\_$154313 & 
        $51.2^{+61.6}_{-11.8}$ & 36.7 & 90\% & This work  \\
 	GW170817 - bright &      
        $68.8^{+17.30}_{-7.63}$ & 12.3 & 30\% & \cite{nicolaou2019impact}  \\
		GW190814 & $78^{+57}_{-13}$ & 35 & 86\%  & \cite{palmese20_sts}\\
        GW190924+GW200202 & $60.33^{+55.79}_{-13.61}$ & 34.7 & 85\%  & This work \\
        GW190814+GW170814 & $77^{+41}_{-22}$ & 31.5 & 77\%  & \cite{palmese20_sts} \\
        8 dark sirens & $79.8^{+19.1}_{-12.4}$ & 15.8 & 39\% & \cite{Palmese_2023}\\
		10 dark sirens & $76.00^{+17.64}_{-13.45}$ & 15.55 & 38\% & This work\\
 	\hline
        \end{tabular}    
 \label{tab:H0results}
 \end{table*}

\begin{figure}
    \centering
    \includegraphics[width=\linewidth]{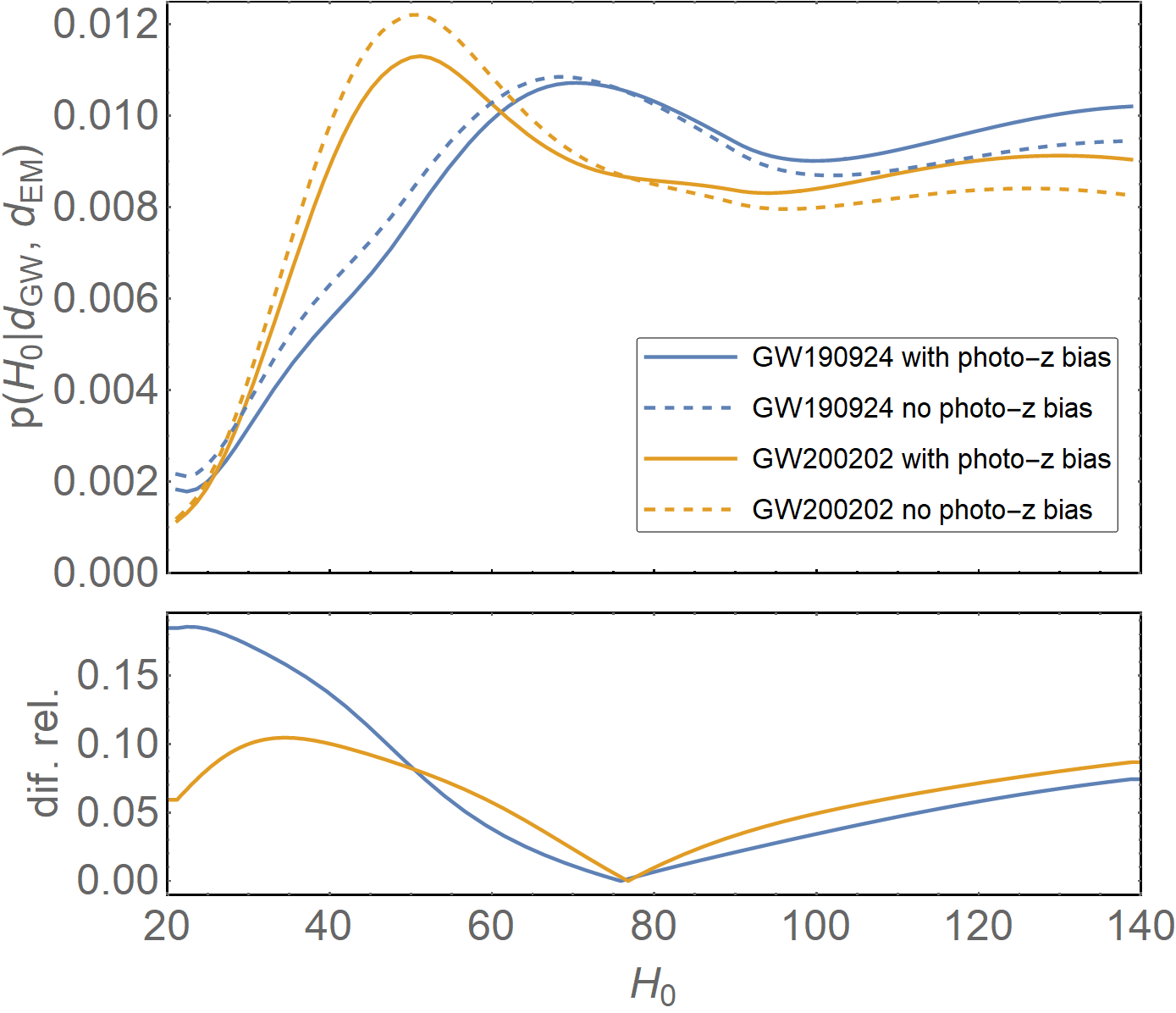}
    \caption{Photo-\textit{z} bias effect on the Hubble constant posterior distributions for GW190924$\_$021846 (blue lines) and GW200202$\_$154313 (yellow lines). Solid curves are the results considering the inclusion of the photo-\textit{z} bias correction, and dashed curves ignore this correction. The sub-plots present the relative difference between these two curves.}
    \label{fig:photozbias}
\end{figure}

Fig.~\ref{fig:photozbias} illustrates the photo-\textit{z} bias effect on the $H_{0}$ posterior distribution for the events GW190924$\_$021846 and GW200202$\_$154313. We see that the effect is a little more significant for GW200202$\_$154313, with a relative difference of $\sim$0.1(0.09) for low(high) $H_0$ values. For the GW190924$\_$021846 dark siren, these values vary from 0.0002 to 0.18. This is different than what was discussed in \citet{palmese20_sts}, which showed that the effect of marginalization over the photo–z bias is minimum for all values of $H_0$. 

\begin{figure}
    \centering
    \includegraphics[width=\linewidth]{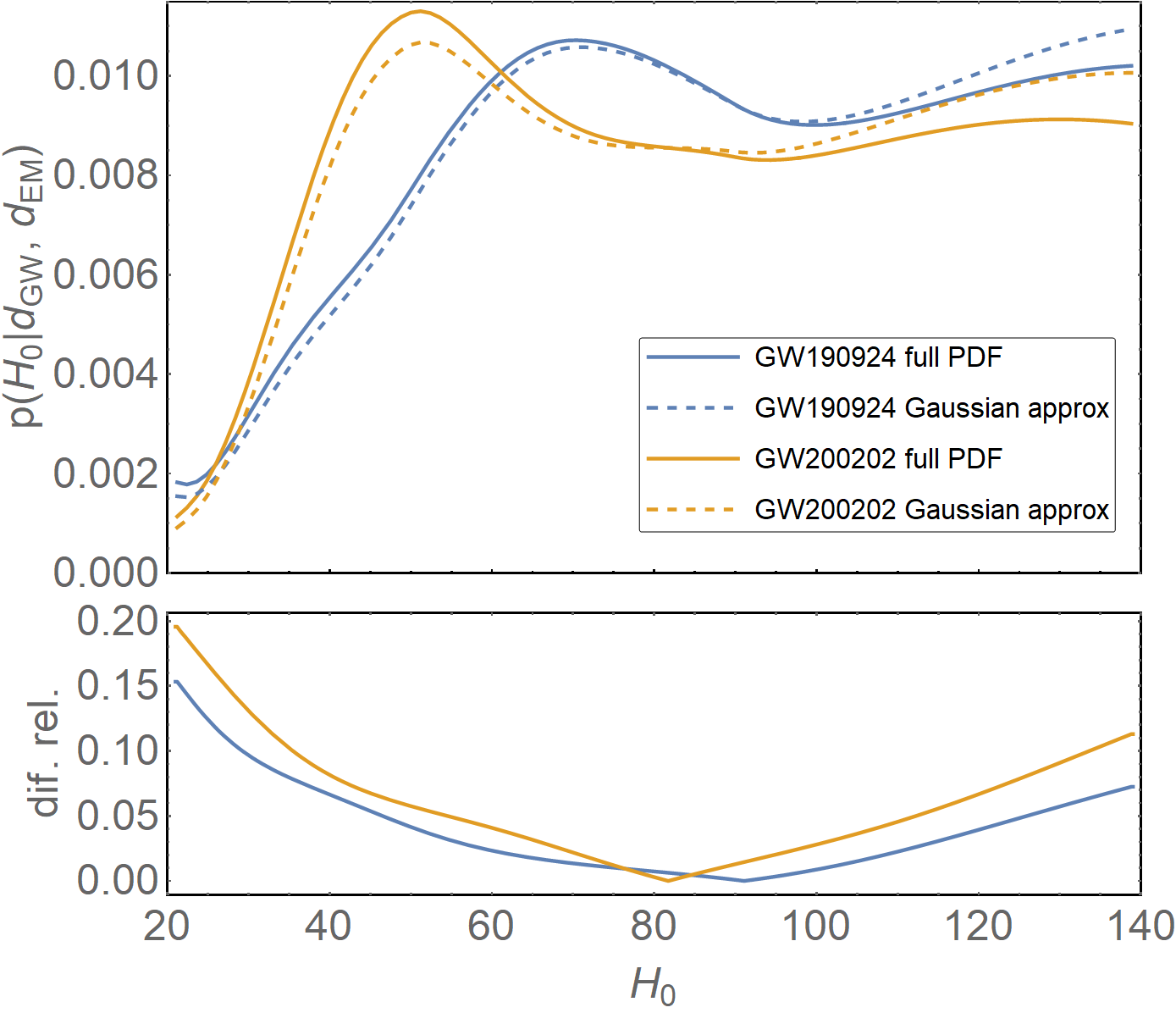}
    \caption{Comparison between the Hubble constant posterior distributions for GW190924$\_$021846 (blue curves) and GW200202$\_$154313 (yellow curves) obtained by using the full galaxies redshift PDFs and a Gaussian approximation. The sub-plots present the relative difference between the curves.}
    \label{fig:zPDF}
\end{figure}

Another correction applied here is the full redshift PDF instead of a Gaussian approximation. The effect of this correction is shown in Fig.~\ref{fig:zPDF}, where it is almost the same for the two dark sirens, with the relative difference varying between 0.0001 and $\sim0.2$ for the entire interval of $H_0$.

In order to understand the impact of dark sirens on the precision of $H_{0}$, we combine our results with the bright siren GW170817 from \citet{nicolaou2019impact}. Fig.~\ref{fig:result} presents the combined $H_0$ posterior. The combination of GW190924$\_$021846, GW200202$\_$154313, and GW170817 gives $H_{0} = 68.84^{+15.51}_{-7.74} \, \rm{km/s/Mpc}$, which is in agreement with the recently presented results in \citet{Abbott_2023}, $H_{0} = 68^{+8}_{-6} \, \rm{km/s/Mpc}$, that used 47 dark sirens (43 BBH, 2 BNS, and 2 NSBH) from the third LIGO–Virgo–KAGRA GW transient catalog \citep{gwtc3} with GLADE+ galaxy catalogs \citep{glade, gladeplus}. This constraint represents a reduction of $\sim$7\% in the 68\% CI of the $H_0$ measurement found with only GW170817. Although our measurement is less precise than \citet{Abbott_2023} (as expected given the smaller number statistics), we note that we expect our result to be less sensitive to the black hole population assumptions. As noted in \citet{Abbott_2023}, these assumptions, and specifically the shape of the mass distribution, strongly dominate the inference on $H_0$. A possible cause of this dependency is the lack of completeness of GLADE+ catalog at the redshifts of interest. By combining these results with the 8 dark sirens from \citet{Palmese_2023}, we find an improvement of 6\% in the precision of the GW170817 measurement. This result highlights the improvement obtained when well-localized GW events at redshifts well covered by galaxy catalogs are incorporated into the analysis.

\section{Conclusions}\label{conclusions}

In this work, we investigate the dark siren method to constrain $H_0$, and present a new measure of $H_0$ provided by two GW events detected by LIGO/Virgo, GW190924$\_$021846 and GW200202$\_$154313, with the redshifts of the potential host galaxies derived using DELVE DR2 data. The estimation of galaxies photo-\textit{z}'s was performed using the deep learning technique Mixture Density Network. Our analyses implement the full redshift PDF of the galaxies instead of the Gaussian approximation. The main result of this study includes the measurement of the Hubble constant of $70.4^{+54.7}_{-15.1}\,\rm{km/s/Mpc}$ and $51.2^{+61.6}_{-11.8}\,\rm{km/s/Mpc}$ for GW190924$\_$021846 and GW200202$\_$154313, respectively, which is consistent with previous measurements of $H_0$. The combination of GW190924$\_$021846, GW200202$\_$154313, together with GW170817 bright siren leads to $H_0 = 68.84^{+15.51}_{-7.74} \, \rm{km/s/Mpc}$, i.e. the addition of the two dark sirens reduces the 68\% CI interval by $\sim$7\%, which is comparable to the $\sim$12\% found in \citet{palmese20_sts} when they add GW190814 and GW170814. This result demonstrates the power of well-localized dark siren events in better constraining the determination of the Hubble constant using deep imaging photometry obtained from surveys performing wide-sky coverage.

In addition, we also present the Hubble constant using only the dark standard siren method. We combine the $H_0$ posteriors found here with the posteriors of the 8 well-localized dark siren events (GW170814, GW190814, GW170608, GW170818, GW190412, S191204r, S200129m and S200311bg) presented by \citet{palmese20_sts} and \citet{Palmese_2023}. The $H_0$ measurement found is $76.00^{+17.64}_{-13.45}$ km/s/Mpc, which has a precision of 20\% and the 68\% CI interval is $\sim$38\% of the prior width. Our result indicates that a sample with ten well-localized dark sirens and a complete galaxy catalog can provide a significant constraint on the Hubble constant that is equivalent to that achieved with a standard siren, providing complementary information to the standard method.

Our results provide an indication of the dark siren potential as a precision cosmological probe. After a period of sensitivity upgrades, over the past few months, the LIGO/Virgo/KAGRA collaboration has returned to operation and is expected to make $\sim90$ detections of mergers per year \citep{2018LRR....21....3A}. With the increase in GW observations and the arrival of deeper and wider surveys, like the forthcoming Vera C. Rubin Legacy Survey of Space and Time (LSST, \citealt{LSST}), it is possible that in the next few years, dark sirens will provide a measure of $H_0$ at the several percentage level \citep{delpozzo}. In this regime, we highlight the need for a more robust analysis, that takes into account potential systematics neglected in the methodology adopted here. Likely, some of the most significant sources of systematics will be the galaxy catalog selection effects, galaxy catalog completeness, the dependence of host galaxy properties on the BBH formation channels, and the use of a Gaussian approximation for the GW likelihood instead of its full asymmetric distribution. In future work, we intend to improve the dark siren methodology in order to consider these corrections.

\section*{Acknowledgements}

For the analysis, we use the Python Programming Language, along with the following package: Astropy \citep{astropy, astropy_v1}, Matplotlib \citep{matplotlib}, NumPy \citep{numpy}, SciPy \citep{scipy}, Tensorflow \citep{tensorflow2015-whitepaper} and ligo.skymap \citep{bayestar, Singer_2016, Singer_supp, Kasliwal_2017}. 

\noindent Gabriel Teixeira and Clecio Bom acknowledges the financial support from CNPq (402577/2022-1)
Clecio Bom acknowledges the financial support from CNPq (316072/2021-4) and from FAPERJ (grants 201.456/2022 and 210.330/2022) and the FINEP contract 01.22.0505.00 (ref. 1891/22). The authors made use of Sci-Mind servers machines developed by the CBPF AI LAB team and would like to thank P. Russano and M. Portes de Albuquerque for all the support in infrastructure matters. C.E.M.-V. is supported by the international Gemini Observatory, a program of NSF’s NOIRLab, which is managed by the Association of Universities for Research in Astronomy (AURA) under a cooperative agreement with the National Science Foundation, on behalf of the Gemini partnership of Argentina, Brazil, Canada, Chile, the Republic of Korea, and the United States of America. GEM acknowledges support from the University of Toronto Arts \& Science Postdoctoral Fellowship program. Time-domain research by D.J.S. is supported by NSF grants AST-1821987, 1813466, 1908972, 2108032, and 2308181, and by the Heising-Simons Foundation under grant \#2020-1864. J.A.C.-B. acknowledges support from FONDECYT Regular N 1220083.

\noindent The DELVE project is partially supported by the NASA Fermi Guest Investigator Program Cycle 9 No. 91201 and NSF awards AST-2108168 and AST-2307126. This work is partially supported by Fermilab LDRD project L2019-011.

\noindent This project used data obtained with the Dark Energy Camera (DECam), which was constructed by the Dark Energy Survey (DES) collaboration. Funding for the DES Projects has been provided by the U.S. Department of Energy, the U.S. National Science Foundation, the Ministry of Science and Education of Spain, the Science and Technology Facilities Council of the United Kingdom, the Higher Education Funding Council for England, the National Center for Supercomputing Applications at the University of Illinois at Urbana-Champaign, the Kavli Institute of Cosmological Physics at the University of Chicago, Center for Cosmology and Astro-Particle Physics at the Ohio State University, the Mitchell Institute for Fundamental Physics and Astronomy at Texas A\&M University, Financiadora de Estudos e Projetos, Fundacao Carlos Chagas Filho de Amparo, Financiadora de Estudos e Projetos, Fundacao Carlos Chagas Filho de Amparo a Pesquisa do Estado do Rio de Janeiro, Conselho Nacional de Desenvolvimento Cientifico e Tecnologico and the Ministerio da Ciencia, Tecnologia e Inovacao, the Deutsche Forschungsgemeinschaft and the Collaborating Institutions in the Dark Energy Survey. The Collaborating Institutions are Argonne National Laboratory, the University of California at Santa Cruz, the University of Cambridge, Centro de Investigaciones Energeticas, Medioambientales y Tecnologicas-Madrid, the University of Chicago, University College London, the DES-Brazil Consortium, the University of Edinburgh, the Eidgenossische Technische Hochschule (ETH) Zurich, Fermi National Accelerator Laboratory, the University of Illinois at Urbana-Champaign, the Institut de Ciencies de l'Espai (IEEC/CSIC), the Institut de Fisica d'Altes Energies, Lawrence Berkeley National Laboratory, the Ludwig-Maximilians Universitat Munchen and the associated Excellence Cluster Universe, the University of Michigan, the National Optical Astronomy Observatory, the University of Nottingham, the Ohio State University, the University of Pennsylvania, the University of Portsmouth, SLAC National Accelerator Laboratory, Stanford University, the University of Sussex, and Texas A\&M University.

\noindent Based on observations at Cerro Tololo Inter-American Observatory, NSF's NOIRLab (NOIRLab Prop. ID 2019A-0305; PI: Alex Drlica-Wagner), which is managed by the Association of Universities for Research in Astronomy (AURA) under a cooperative agreement with the National Science Foundation.

\noindent BASS is a key project of the Telescope Access Program (TAP), which has been funded by the National Astronomical Observatories of China, the Chinese Academy of Sciences (the Strategic Priority Research Program "The Emergence of Cosmological Structures" Grant \# XDB09000000), and the Special Fund for Astronomy from the Ministry of Finance. The BASS is also supported by the External Cooperation Program of Chinese Academy of Sciences (Grant \# 114A11KYSB20160057), and Chinese National Natural Science Foundation (Grant \# 11433005).

\noindent This manuscript has been authored by Fermi Research Alliance, LLC, under contract No. DE-AC02-07CH11359 with the US Department of Energy, Office of Science, Office of High Energy Physics. The United States Government retains and the publisher, by accepting the article for publication, acknowledges that the United States Government retains a non-exclusive, paid-up, irrevocable, worldwide license to publish or reproduce the published form of this manuscript, or allow others to do so, for United States Government purposes.

\section*{Data Availability}
 
The data underlying this article will be shared on reasonable request to the corresponding author.



\bibliographystyle{mnras}
\bibliography{references} 








\bsp	
\label{lastpage}
\end{document}